\documentclass[twoside]{ilcws07}
\usepackage[latin1]{inputenc}
\usepackage[dvips]{graphicx,epsfig,color}
\usepackage{wrapfig,rotating}
\usepackage{amssymb,amsmath,array}

\begin{document}

\noindent
\begin{minipage}[t]{.2\linewidth}
\leavevmode
 \hspace*{-.8cm}
\includegraphics{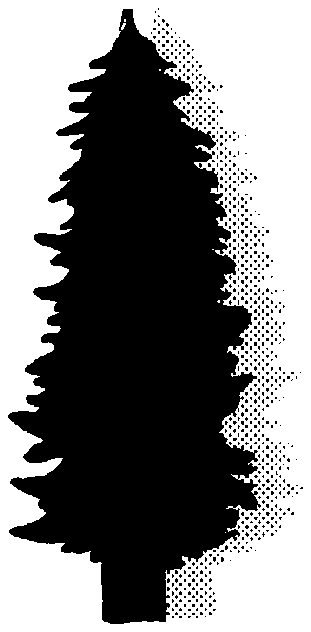}
\end{minipage} \hfill
\begin{minipage}[b]{.45\linewidth}
\rightline{SCIPP 07/14}
\rightline{September 2007}
\vspace{3cm}
\end{minipage}
\vskip1.5cm

\thispagestyle{empty}

\begin{center}
{\large\bf SIMULATION OF AN ALL-SILICON TRACKER}\\[2pc]

Chris Meyer, Tyler Rice, Lori Stevens, Bruce A. Schumm \\

\vskip0.2cm

University of California at Santa Cruz and the Santa Cruz Institute for Particle
Physics \\

1156 High Street, Santa Cruz, California 95062, USA \\

\end{center}
\vskip1cm

We present recent improvements in the performance of the
reconstruction of non-prompt track in the SiD Detector
Concept Design, including initial results on the effect
of longitudinal segmentation in the SiD tracker. We also
describe a generic tracking validation package developed
at SCIPP.

\vfill
\begin{center}
{\small Talk presented at the 2007 International Linear Collider
Workshop, DESY, Hamburg, Germany, May 30 -- June 3, 2007}
\end{center}
\vfill
\clearpage

\title{
Simulation of an All-Silicon Tracker} 
\author{Chris Meyer, Tyler Rice, Lori Stevens, Bruce A. Schumm
\vspace{.3cm}\\
University of California at Santa Cruz and the Santa Cruz Institute for Particle 
Physics \\
1156 High Street, Santa Cruz, California 95062, USA
}

\maketitle

\begin{abstract}
We present recent improvements in the performance of the 
reconstruction of non-prompt track in the SiD Detector
Concept Design, including initial results on the effect
of longitudinal segmentation in the SiD tracker. We also
describe a generic tracking validation package developed
at SCIPP.
\end{abstract}

The SiD Detector Concept offers a number of potential
advantages in the reconstruction of ILC collisions, but
to be confident of the quality of this reconstruction,
several of the SiD's innovative features need to be studied
via realistic simulation. In particular, it must be
demonstrated that the use
of a limited number (five in the baseline design)
of precise silicon layers for track reconstruction
is sufficient to exploit the physics potential of
the ILC machine. While there is little question 
that such a device can provide superior transverse
momentum resolution, the capability of the design to 
recognize and
reconstruct tracks  -- particularly those that originate
outside the first layer or two of the vertex
detector -- is less clear.

To explore the performance of the tracker under
realistic settings, we have developed a tracking
performance package that evaluates the tracking
efficiency and reconstruction accuracy of tracking
algorithms. This package has been developed as
a stand-alone C++ package, and thus is fully versatile,
being easily applied to any ILC detector concept
within any reconstruction framework (in fact, it
is generally applicable to any cylindrical 
geometry track reconstruction package). In the 
interest of space, we do not present further 
details of this package here; sample output is
to be found in reference ~\cite{schumm}. We encourage
interested groups to contact us for details.

One of the novel aspects of the baseline SiD design is
that it proposes to do tracking reconstruction
with only ten tracking layers, of which five are
concentrated around the beampipe in a pixelated
vertex detector. While previous studies ~\cite{sinev,schumm}
have suggested that such an approach can
be sufficiently efficient for prompt tracks, the
reconstruction algorithm used in these prior studies
was unable to reconstruct tracks originating 
outside of the first layer of the vertex detector.
Over the past year, the SCIPP ILC simulation group
has explored the capability of the SiD tracker to
reconstruct such non-prompt tracks, and how that
capability depends upon the longitudinal segmentation
of the tracker. The results presented here are
somewhat updated relative to the results presented
at the workshop in May 2007.

The SCIPP group's work on non-prompt track reconstruction
has been based on refining and extending AxialBarrelTracker,
an algorithm originally written by Tim Nelson (SLAC) to
reconstruct SiD tracks in the absence of the vertex
detector. This algorithm
works inward from the outside of the SiD tracker, beginning
with three-hit seeds that lie on circles in the plane
transverse to the beam line
that miss the collision point by no more than 1 centimeter.
To search for non-prompt tracks, the SCIPP group relaxed the
miss-distance requirement to 10 centimeters, finding that, once
the hits from prompt tracks were removed, the number of
seeds remained tractable with the relaxed DOCA constraint.

The group explored the nature of the hits remaining after
the hits from prompt tracks were removed. Roughly 5\% of
hits were due to tracks that went through three or more tracking layers 
and then exited the tracker. Approximately 45\% of the
hits appeared to be coming from tracks that looped through
the tracker, striking each tracking layer a number of times. Roughly
35\% of hits were due to material interactions of prompt tracks.
The remaining 15\% were hits from tracks with momentum too low
to reconstruct.

In this light, a set of `findable' non-prompt particles was identified
by requiring that the underlying (`Monte-Carlo Truth')
tracks lie within $|\cos \theta| < 0.8$, have a radius of origin 
between 2 and 40 cm and a path length in the tracker of of at least 50cm, 
not arise via back-scatter off of the calorimeter, and have a
transverse momentum of no less than 0.75 GeV/c. In a sample of 137
Z-pole $b {\overline b}$ events with thrust greater than 0.94 and
a thrust axis with $|\cos \theta _{thrust}| < 0.5$, these selection
requirements identified 304 findable non-prompt particles. This set
of findable particles represents approximately 5\% of the number of
findable particles that would be identified if there were no restriction
on the radius of origin (in other words, approximately 5\% of all
tracks are non-prompt). 

Tracks found by AxialBarrelTracker were accepted provided they
were comprised of at least four hits, had a reconstructed
transverse momentum of at least 0.75 GeV/c, and a reconstructed
distance of closest approach in the plane perpendicular to the
beam of no more than 10 cm. 
The results that follow were achieved under the
assumption that the tracker was composed of two unsegmented
halves: one extending to positive value of $z$ and the other to
negative values; only tracks for which all hits had the same
sign $z$ coordinate were accepted.

Findable particles were deemed `found' provided they were associated
with accepted tracks that had at least four hits caused by
the findable particle under consideration. No more than one accepted track
was permitted to be associated with each findable particle. 
Any accepted track not associated with a findable particle was 
deemed `fake'. 

With these criteria, 131 (43\%) of findable non-prompt particles were found
with 5 hits, with only one fake five-hit track. Another 100 non-prompt
tracks were found with 4 hits; however, these were accompanied by an additional 
270 four-hit fake tracks, rendering four-hit tracks too impure for use. The
remaining findable particles (73, or 24\% of the sample) had no associated
accepted track.

Upon examination, it was discovered that AxialBarrelTracker was often
being confused by three-hit seeds for which not all of the hits came
from the same underlying particle. Thus, to improve the efficiency
as well as reduce the number of fake tracks, we added to AxialBarrelTracker a
requirement that all the hits on the seed lie within an azimuthal
slice of width $\pi / 2$. This requirement was also applied to the larger
set of hits as additional hits were added to the seed. 

After application
of this azimuthal restriction, the sample of 304 findable particles
were reconstructed as follows. 145 (48\%) were reconstructed with 
five hits, 112 (37\%) were reconstructed with four hits, and 47 (15\%)
had no associated accepted track. The number of five-hit fake tracks 
remained unchanged at one, while the number of four hit fake tracks
was reduced from 270 to 157.
It should be point out that, of the 304 findable particles, only 166
left one and only one hit in each of the five layers; for this set of
tracks, the reconstruction efficiency with the azimuthal restriction approached 85\%.

\begin{figure}
\centerline{\includegraphics[width=1.0\columnwidth]{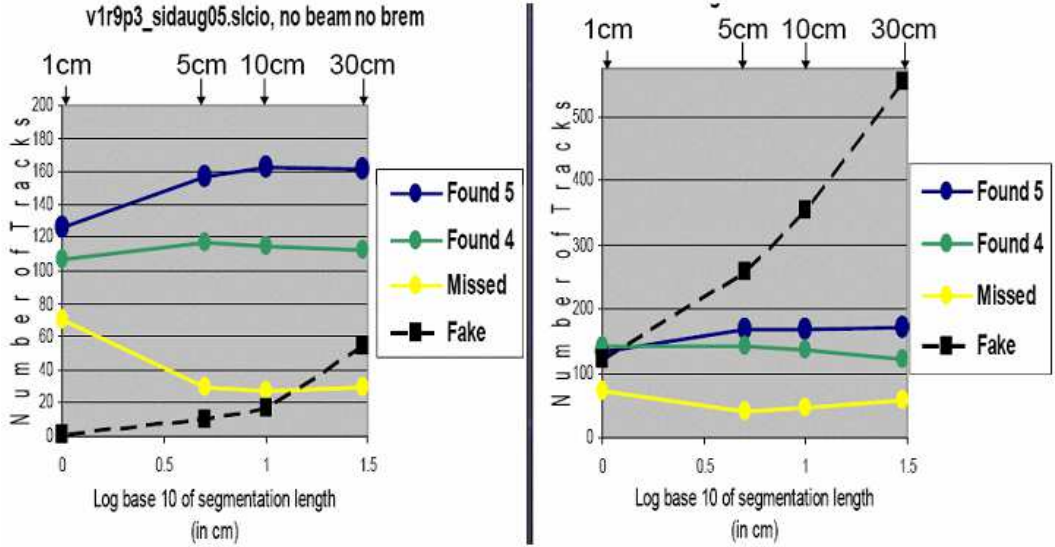}}
\vspace{-27mm}
\caption{Performance of AxialBarrelTracker as a function of
the $z$ length of the tracker modules for Z pole 
$b {\overline b}$ events (left) and $uds$ events at
$E_{cm} = 500$ GeV. The trajectories correspond to the
numbers of findable particles found with five hits and
four hits, the number of findable particles with no
associated track, and the number of fake tracks.}\label{Fig:zsg}
\end{figure}

Thus, it appears that non-prompt particles leaving hits in all five central
tracking layers can be reconstructed with reasonable efficiency and
good purity, but that more work needs to be done to reconstruct 
particles leaving only four hits (the majority of four-hit particles
originate outside of the first central tracking layer, rather than originating 
at smaller radius but curling up or leaving the detector before reaching the fifth
tracking layer). To this end, the group developed 
code that takes advantage of longitudinal segmentation, testing for
the consistency of the pattern of struck modules with the hypothesis
that the pattern was produced by a single particle. The performance
of AxialBarrelTracker as a function of the $z$ length of the tracker
modules is shown in Figure \ref{Fig:zsg}, for two data samples: 
$b {\overline b}$ events at the Z pole and $uds$ events at 
$E_{cm} = 500$ GeV. As stated above, the $b {\overline b}$ events 
contained 304 findable non-prompt tracks; the $E_{cm} = 500$ GeV
events contained 352 findable non-prompt tracks. It is seen that
10 cm segmentation (the SiD baseline) can be very helpful for
intermediate energy ($\sim$50 GeV) jets, but does not appear to 
make too a qualitative difference for high energy ($\sim$250 GeV) jets.

In summary, the SCIPP simulation group has optimized Tim Nelson's
AxialBarrelTracker routine to find non-prompt tracks. The efficiency
and purity for non-prompt tracks hitting all five tracking layers is
good, but without $z$ segmentation, it seems difficult to reconstruct
tracks that hit four or fewer layers. The inclusion of $z$ segmentation
can provide a substantial benefit by reducing the number of fake
four-hit tracks. However, the degree of segmentation needed to 
reduce the fake-track contribution enough to make four-hit tracks
usable depends upon the physics being studied. For low-energy (~45
GeV) jets, the proposed 10cm segmentation of the SiD baseline
may be sufficient. For high-energy (250 GeV) jets, however, 
the current reconstruction seems to require segmentation 
on the order of 1cm or less to recover four-hit tracks.
The SCIPP simulation group continues to explore the capability
of the SiD baseline tracker to reconstruct non-prompt tracks, and
is in the process of implementing the GARFIELD tracker ~\cite{garfield},
which uses minimum-ionizing calorimeter stubs to seed tracks,
as an additional layer in the SiD reconstruction.


\begin{footnotesize}


\end{footnotesize}


\end{document}